\documentclass[useAMS,usenatbib,usegraphicx]{mn2e}

\def\aj{AJ}%
\def\apj{ApJ}%
\def\apjl{ApJ}%
\def\apjs{ApJS}%
\def\mnras{MNRAS}%
\def\nat{Nature}%
\def\iaucirc{IAU~Circ.}%
\title{A single radio-emitting nucleus in the dual AGN candidate NGC~5515}

\author[K.\'E. Gab\'anyi, S. Frey, T. Xiao, Z. Paragi, T. An, E. Kun and L.\'A. Gergely]{K.\'E. Gab\'anyi$^{1,2}$\thanks{E-mail:
gabanyi@konkoly.hu}, S. Frey$^{3}$, T. Xiao$^{4,5}\thanks{LAMOST fellow}$, Z. Paragi$^{6}$, T. An$^{5,7}$, E. Kun$^{1}$ and L.\'A. Gergely$^{1,8}$\\
$^{1}$Departments of Theoretical and Experimental Physics, University of Szeged, D\'om t\'er 9, H-6720 Szeged, Hungary\\
$^{2}$Konkoly Observatory, MTA Research Centre for Astronomy and Earth Sciences, P.O. Box 67, H-1525 Budapest, Hungary\\
$^{3}$F\"OMI Satellite Geodetic Observatory, P.O. Box 585, H-1592 Budapest, Hungary\\
$^{4}$Partner Group of Max Planck Institute for Astrophysics and Key Laboratory for Research in Galaxies and Cosmology of Chinese Academy \\ of Sciences, P.R. China \\
$^{5}$Shanghai Astronomical Observatory, Chinese Academy of Sciences, 80 Nandan Road, 200030 Shanghai, P.R. China\\
$^{6}$Joint Institute for VLBI in Europe, Postbus 2, 7990 AA Dwingeloo, The Netherlands\\
$^{7}$Key Laboratory of Radio Astronomy, Chinese Academy of Sciences, P.R. China\\
$^{8}$Department of Physics, Tokyo University of Science, Shinjuku-ku, Tokyo, Japan}

\begin{document}

\date{Accepted 2014 June 16.  Received 2014 June 12; in original form 2014 May 14}

\pagerange{\pageref{firstpage}--\pageref{lastpage}} \pubyear{2014}

\maketitle

\label{firstpage}

\begin{abstract}
The Seyfert galaxy NGC~5515 has double-peaked narrow-line emission in its optical spectrum, and it has been suggested that this could indicate that it has two active nuclei.
We observed the source with high resolution Very Long Baseline Interferometry (VLBI) at two radio frequencies, reduced archival Very Large Array data, and re-analysed its optical spectrum. We detected a single, compact radio source at the position of NGC~5515, with no additional radio emission in its vicinity. The optical spectrum of the source shows that the blue and red components of the double-peaked lines have very similar characteristics. While we cannot rule out unambiguously that NGC~5515 harbours a dual AGN, the assumption of a single AGN provides a more plausible explanation for the radio observations and the optical spectrum.
\end{abstract}


\section{Introduction}
It is widely accepted that most massive galaxies harbour supermassive black holes (SMBHs) in their centres \citep{salpeter,lbell}. In hierarchical structure formation models, interactions and mergers between galaxies play an important role in their evolution and consequently in the growth of their central SMBHs. Thus, it is expected that a particular phase in the merging process, namely systems with dual SMBHs exist in the Universe. %

In such systems, one or both of the SMBHs may be active; several studies suggest that the merging process can cause enhanced accretion onto the central SMBHs and thus initiate ``activity'' \citep[e.g.][]{Dima2005}. High-resolution particle hydrodynamical simulations \citep{VW} suggest that simultaneous activity is mostly expected at the late phases of mergers, at or below 10 kpc-scale separations. Therefore dual active galactic nuclei (AGN, with two active SMBHs in a merger system) are expected to be observed. So far, only a few convincing cases of dual AGN are known \citep[e.g.][]{Komossa, rodriguez06, Bondi, Liu_disc, Shen2011, Fu2012}. In some cases \citep[e.g.][]{Liu_disc}, high spatial-resolution optical photometry and spectroscopy or high resolution Very Long Baseline Interferometry (VLBI) provided the unambiguous evidence of the dual system.

Originally, it was thought that the presence of double-peaked narrow optical emission lines is indicative of the existence of dual AGN, as these lines may originate from the two distinct narrow line regions (NLR) of the two AGN \citep{nlr_wang}. However, several studies already in the eighties \citep[][]{heckman81, heckman84} showed that double-peaked narrow emission lines can arise due to peculiar kinematics and jet--cloud interaction in a single NLR. As of now, there is no known observational approach which could be used to select a large sample of compelling dual AGN candidates. 
Therefore, it is crucial to check with independent methods whether the candidate sources are indeed dual systems. 

Recently, \cite{Beni13a} reported the serendipitous discovery of two new dual AGN candidates. They studied a sample of ten close-by, intermediate-type Seyfert galaxies chosen from the Sloan Digital Sky Survey \citep[SDSS,][]{sdssdr4} database. The intermediate-type Seyfert galaxies (spanning from Sy 1.2 to Sy 1.9) belong to Seyfert 1 galaxies as they show broad emission lines, however with decreasing intensity \citep{o81}. The role of this kind of AGN within the unified scheme is not clear. Surprisingly, half of the sample studied by \cite{Beni13a} showed narrow double-peaked emission lines. Based upon the line ratios (O{\sc iii}/H$\beta$, N{\sc ii}/H$\alpha$, S{\sc ii}/H$\alpha$ and O{\sc i}/H$\alpha$), \cite{Beni13a} concluded that among the five double-peaked narrow-line emitters, two (NGC\,5515 and Mrk\,1469) are good candidates for being dual AGNs. 

Both of the sources are radio emitters. We conducted high-resolution radio imaging observations of the brighter radio source, NGC~5515, with the European VLBI Network (EVN). The best resolution achieved was $\sim 2$\,milli-arcseconds (mas), which corresponds to a projected linear distance of $1.04$\,pc in the rest frame of the source, at a redshift of $z$=0.0257, assuming a flat $\Lambda$CDM cosmological model with $H_0$=70~km\,s$^{-1}$\,Mpc$^{-1}$, $\Omega_{\rm m}$=0.27, and $\Omega_\Lambda$=0.73 \citep{Wrig06}. In this model, the luminosity distance of NGC~5515 is $D_{\rm L}$=112.3~Mpc.

Our observations of NGC~5515 and the details of data reduction are given in Section~\ref{observations}, including the analysis of archival multi-frequency data obtained with the US National Radio Astronomy Observatory (NRAO) Very Large Array (VLA). The results are presented in Section~\ref{results} and further discussed, together with the related optical data,  in Section~\ref{discussion}. The summary is given in Section~\ref{sum}.

\section{Observations and data reduction}
\label{observations}

\subsection{VLBI observations}

\begin{figure*}
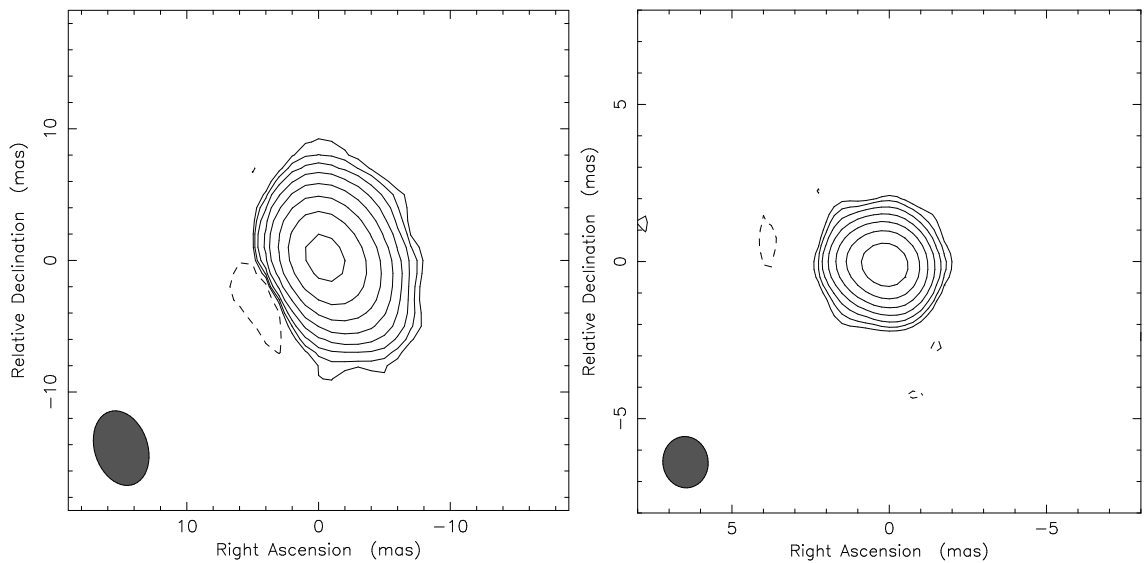

\centering
  \includegraphics[bb=68 176 508 619, height=75mm, angle=270, clip=]{NGC5515-Lband.ps}
  \includegraphics[bb=68 169 522 626, height=75mm, angle=270, clip=]{NGC5515-Cband.ps}
  \caption{
EVN images of NGC~5515 at 1.7~GHz {\it (left)} and 5~GHz  {\it (right)}. The lowest negative and positive contours are drawn at around 3$\sigma$ image noise levels. Further positive contour levels increase by a factor of 2. The restoring beam (full width at half maximum, FWHM) is indicated with ellipses in the lower-left corners. The coordinates are related to the brightness peak at right ascension $14^{\rm h}12^{\rm m}38\fs15423$ and declination $39\degr18\arcmin36\farcs8162$. In the 1.7-GHz image {\it (left)}, the lowest contour level is $\pm60$~$\mu$Jy~beam$^{-1}$, the peak brightness is 9.72~mJy~beam$^{-1}$, the restoring beam is 5.82~mas~$\times$~4.03~mas with a major axis position angle $18\degr$. In the 5-GHz image {\it (right)}, the lowest contour level is $\pm130$~$\mu$Jy~beam$^{-1}$, the peak brightness is 13.24~mJy~beam$^{-1}$, the restoring beam is 1.64~mas~$\times$~1.44~mas with a major axis position angle $7\degr$.}
  \label{image}
\end{figure*}

We observed the nucleus of NGC~5515 with the EVN at 1.7 and 5~GHz frequencies (project codes EG070A and EG070B). The observations were conducted in e-VLBI mode \citep{Szom08} where the participating radio telescopes are connected to the central EVN data processor at the Joint Institute for VLBI in Europe (JIVE, Dwingeloo, the Netherlands) via optical fibre networks, to allow correlation in real time. The maximum data 
transmission rate per station was 1024~Mbit~s$^{-1}$, resulting in a total bandwidth of 128~MHz in both left and right circular polarizations, using 2-bit sampling. Both experiments lasted for 4~h. Their dates (2013 April 16 at 1.7~GHz and 2013 June 18 at 5~GHz) were chosen close to each other to minimize the effects of potential long-term source variability. At 1.7~GHz, the following 7 radio telescopes provided useful data: Effelsberg (Germany), Medicina (Italy), Onsala (Sweden), Toru\'n (Poland), Hartebeesthoek (South Africa), Sheshan (China), and the phased array of the Westerbork Synthesis Radio Telescope (WSRT, the Netherlands). At 5~GHz, the successfully participating 7 radio telescopes were Effelsberg, Medicina, Toru\'n, Hartebeesthoek, Sheshan, the WSRT, and Yebes (Spain).   

NGC~5515 was observed in phase-reference mode \citep[e.g.,][]{Beas95}. By regularly changing the pointing direction of the telescopes between the target source and a sufficiently bright and compact nearby calibrator, the coherent integration time on the target can be extended and thus the imaging sensitivity improved. The quasar J1419+3821 was selected as the phase-reference calibrator, at $1\fdg68$ angular separation from NGC~5515. The calibrator is one of the defining sources of the current 2nd realization of the International Celestial Reference Frame \citep[ICFR2,][]{Fey09}, at right ascension $\alpha_0$=$14^{\rm h}19^{\rm m}46\fs6137607$ and $\delta_0$=$38\degr21\arcmin48\farcs475093$, with a formal uncertainty of 0.04~mas in each coordinate. Within each target--reference cycle of $\sim$5~min, NGC~5515 was observed for 3.3~min, accumulating a total on-source integration time of $\sim$2.4~h at both frequencies. In an ideal case with no loss of data, the expected image thermal noise level was 14~$\mu$Jy~beam$^{-1}$ and 16~$\mu$Jy~beam$^{-1}$ at 1.7 and 5~GHz, respectively\footnote{EVN Calculator: http://www.evlbi.org/cgi-bin/EVNcalc}. The noise levels achieved in practice depend on various factors e.g. downtimes, actual system temperatures, data rate limitations, and radio-frequency interference at the telescope sites.
                                 
The NRAO Astronomical Image Processing System \citep[{\sc AIPS},][]{Grei03} was used for the data calibration in a standard way \citep[e.g.][]{Diam95}. The visibility amplitudes were calibrated using system temperatures and antenna gains measured at the telescope sites. Fringe-fitting was performed for the calibrator source J1419+3821. The calibrator data were then exported to the {\sc Difmap} package \citep{Shep94} for imaging. The conventional hybrid mapping procedure involving several iterations of CLEANing \citep{Hogb74} and phase (then amplitude) self-calibration resulted in an image and a brightness distribution model for the practically unresolved calibrator. Overall antenna gain correction factors (typically less than 10 per cent) were determined in {\sc Difmap} and applied to the visibility amplitudes in {\sc AIPS}. Fringe-fitting was repeated for the phase-reference calibrator in {\sc AIPS}, now taking its CLEAN component model into account in order to compensate for small residual phases resulting from its structure. The solutions obtained were interpolated and applied to the NGC~5515 data. The calibrated and phase-referenced visibility data of NGC~5515 were imaged in {\sc Difmap}.

Because the a priori position of the central radio source in NGC~5515 used for correlation was accurate to only $\sim$$0\farcs5$, the phase centre was shifted to the location of the brightness peak. After obtaining an initial CLEAN component model, a phase-only self-calibration was performed for 15-min solution intervals, to correct for long-period phase variations at the antennas. The source appeared the most resolved on the baselines to Sheshan, therefore data from this telescope were fixed before another CLEANing and phase self-calibration was performed, now with 5-min solution intervals. Finally, only the phases at the two most sensitive antennas, Effelsberg and the WSRT were allowed to vary when the self-calibration solution interval was set to zero. 
No amplitude self-calibration was done on the target source. The total intensity images restored with the final CLEAN component models are displayed in Fig.~\ref{image}. The weights of data points were made inversely proportional to the amplitude errors, by setting {\it uvweight 0,$-$1} (natural weighting) in {\sc Difmap}. 

\subsection{Archival VLA data}

We also reduced VLA observations of NGC~5515 found in the NRAO archive. The source was observed in the most extended A configuration at L, C, and X bands (at 1.5, 5, and 8 GHz) in snapshot mode on 1991 Aug 24, 1992 Oct 20, and 1995 Aug 14, respectively. (The project codes were: AC301, AF233 and AM484). Additionally, the source was observed in snapshot mode at L band in B configuration on 1993 Apr 26 (project id.: AT149) as well. The on-source integration times were less than 5 min in all cases.

NGC~5515 was detected as a point source in all three bands with the following flux densities: in A-configuration $S_\mathrm{L}^\mathrm{A}=(16\pm 1)$\,mJy, $S_\mathrm{C}=(16 \pm 1)$\,mJy, and $S_\mathrm{X}=(29\pm 1)$\,mJy; and in B-configuration $S_\mathrm{L}^\mathrm{B}=(26\pm 3)$\,mJy.

The source was also observed with the VLA at L band in the NRAO VLA Sky Survey \citep[NVSS,][]{nvss} and the Faint Images of the Radio Sky at Twenty-Centimeters \citep[FIRST,][]{first} surveys. The NVSS observation was conducted in D configuration in April 1995, the flux density is $(28.8\pm1)$\,mJy. The FIRST observation was conducted in B configuration in 1994, the flux density is $(19.19\pm0.14)$\,mJy.

\section{Results}
\label{results}

Our EVN images in Fig.~\ref{image} show a single compact mas-scale radio source. Based on the slight asymmetry of the contours, there is a hint of a somewhat more extended structure in about the east--west direction. This notion is supported by the fact that the interferometer phases on the baselines from the European antennas to Hartebeesthoek (i.e. north--south direction) appeared less noisy than on the baselines to Sheshan (east--west direction, at about the same baseline length), therefore the source seems more resolved in the latter direction. 

To quantitatively characterize the brightness distribution of NGC~5515, we fitted Gaussian model components directly to the self-calibrated VLBI visibility data in {\sc Difmap}. The parameters of the best-fitting elliptical Gaussian model components are given in Table~\ref{modelfit}. The statistical errors are estimated according to \citet{Foma99}, assuming additional 5\% flux density calibration uncertainties. The sizes of the fitted model components exceed the values obtained for the minimum resolvable angular size \citep[e.g.][]{Kova05} in our experiments. Consistently with our previous remark on the possible extension, the major axes of the Gaussians at both frequencies closely align with the east--west direction (i.e. position angle 90\degr; the position angles are measured from north through east). Notably, these are almost coincident with the major axis direction of the disk and the pseudobulge of the host galaxy \citep[103--104\degr,][]{Beni13b}.

\begin{table*}
  \begin{center} 
  \caption[]{Parameters of the central elliptical Gaussian model components for NGC~5515, and the inferred brightness temperatures.}
  \label{modelfit}
\begin{tabular}{cccccc}        
\hline                 
Frequency   & Flux density   & \multicolumn{2}{c}{Component axes (FWHM)} & Major axis  & Brightness temperature \\
$\nu$ (GHz) & $S$ (mJy)      & $\vartheta_1$ (mas)  & $\vartheta_2$ (mas) & position angle (\degr) & $T_{\rm B}$ ($10^9$~K)  \\
\hline                       
1.7         & 12.07$\pm$0.61 &  2.241$\pm$0.003 & 1.136$\pm$0.003 & 107 & 2.14$\pm$0.12  \\ 
5           & 16.52$\pm$0.83 &  0.940$\pm$0.001  &  0.120$\pm$0.001  &  86 & 7.36$\pm$0.44  \\ 
\hline   
\end{tabular}
\end{center}
\end{table*}

Based on the fitted VLBI component flux densities, the two-point spectral index is $\alpha_{1.7}^{5}=0.29$ (where the spectral index is defined as $S\propto\nu^{\alpha}$). This indicates a slightly inverted radio spectrum, unless the source was strongly variable between the two observing epochs separated by $\sim$2 months. 

At 5 GHz, the WSRT data taken parallel with the EVN observation were also analyzed. The recovered flux density is $19.6\pm 0.09$\,mJy, close to the value obtained by EVN (Table \ref{modelfit}). Thus, the source is dominated by the emission from the compact core, the large-scale structure resolved out by EVN observation accounts for $\sim 3$\,mJy ($\sim15$ per cent).

According to the archival VLA A-configuration observations, the source is significantly brighter at 8\,GHz than at lower frequencies. However, the spectral index is also compatible with a flat spectral shape as well as an inverted spectral shape ($\alpha=0.39 \pm 0.31$). Moreover, there is a large temporal gap between the observations (three years), therefore intrinsic source variability may hinder the estimation of the spectral index of the source. Indeed, the variability is apparent, when comparing observations performed at the same resolution and frequency (VLA B-configuration at 1.4 GHz). The source was significantly brighter (more than $30$ per cent) in April 1993 than in the FIRST survey observation conducted in 1994. 

Even though the spectral index cannot be calculated reliably from the archival VLA observations, the measurements agree with the source having most likely flat or slightly inverted spectrum and thus in accord with the spectral index determined from our EVN measurements. The displayed variability and the spectral shape is consistent with the radio emission coming from a compact partially self-absorbed synchrotron source \citep{bk}.

The astrometric position of the 5-GHz radio brightness peak in NGC~5515 (right ascension $\alpha$=$14^{\rm h}12^{\rm m}38\fs15423$ and declination $\delta$=$39\degr18\arcmin36\farcs8162$) was derived using the {\sc MAXFIT} verb in {\sc AIPS}. We estimate that each coordinate is accurate to within 1~mas. The sources of the positional error are the thermal noise of the interferometer phases, the error of the phase-reference calibrator position, and the systematic error of phase-referencing observations mainly originating from the ionospheric and tropospheric fluctuations. In our case, the latter is by far the most dominant error component. The position of the brightness peak at 1.7~GHz coincides with the 5-GHz position well within the uncertainties.   

A large window of $8\farcs2 \times 8\farcs2$ around the brightness peak, within the undistorted field of view, \footnote{The undistorted field of view is defined as an area where the expected brightness loss for a point source is less than 10 per cent with respect to the pointing centre.} was checked for additional radio emission. This size corresponds to a region of 4.25~kpc~$\times$~4.25~kpc at the distance of NGC~5515. No other compact radio component was found above the $\sim$6$\sigma$ image noise level of 0.13~mJy~beam$^{-1}$ at 1.7~GHz. We also checked the field of view of the archival VLA data for possible radio sources. According to the optical SDSS image, the size of the NGC\,5515 galaxy is roughly $80\arcsec \times 60\arcsec$. Within this range we did not detect any additional radio emitting source in the VLA images above $\sim 6\sigma$ image noise level ($1$-$2$\,mJy \,beam$^{-1}$).

\section{Discussion}
\label{discussion}

We calculated the rest-frame brightness temperature of the radio source in NGC~5515,
\begin{equation}
T_{\rm B} = 1.22\times 10^{12} \frac{S}{\vartheta_1 \vartheta_2 \nu^2} (1+z) \,\,{\rm K},
\end{equation}
($S$ is given in Jy, $\vartheta_1$ and $\vartheta_2$ in mas, and $\nu$ in GHz) using the Gaussian model parameters listed in Table~\ref{modelfit}. The values obtained (higher than $10^9$~K, see Table~\ref{modelfit}) clearly prove the AGN-related non-thermal synchrotron origin of the radio emission since the brightness temperatures for thermally-dominated ``normal'' galaxies (i.e. the ones with no central active nucleus) do not exceed $\sim$$10^{5}$~K \citep[e.g.][]{Cond92}. 

Our non-detection of any additional compact source in the field of view implies an upper limit of the radio power $P\la 2 \times 10^{20}$~W\,Hz$^{-1}$. According to e.g. \cite{Kewl00} and \cite{Midd11}, AGN have high-luminosity cores, with power exceeding $ 2 \times 10^{21}\mathrm{\,W\,Hz}^{-1}$, therefore we can rule out the existence of another radio-emitting AGN in the field of view. 
We can thus conclude that there is no dual radio-emitting AGN in the centre of NGC~5515. 

The coordinates of the detected radio source agree with the position of the optical galaxy in SDSS DR9 within the errors. (The distance between the optical and radio positions is $\sim 0\farcs22$.) According to the SDSS, there is an optical source $\sim9\arcsec$ away from the radio position, SDSS J141237.38+391835.2, at a similar redshift ($z=0.026$). We checked the L-band and C-band EVN maps at the position of this source and we did not find any radio source down to a brightness limit of $150\mathrm{\,}\mu\mathrm{Jy\,beam}^{-1}$ and $200\mathrm{\,}\mu\mathrm{Jy\,beam}^{-1}$ in L band and C band, respectively. These brightness limits were calculated by taking into account the bandwidth and time-average smearing effects \citep{bs}.
There is also an X-ray source at a distance of $\sim 7\arcsec$ from the radio position in the ROSAT All-Sky Survey Faint Source Catalog \citep[RASS-FSC,][]{rosat_faint}. Since its positional uncertainty is $11\arcsec$, it can be related to either of the two optical sources. However, given that our radio observations indicate an AGN in the nucleus of NGC\,5515, the X-ray emission is most likely associated with that object. 

We re-analyzed the SDSS spectrum of NGC~5515. Similarly to \cite{Beni13a}, we found that the narrow H\,$\beta$, H\,$\alpha$, O\,{\sc iii}, N\,{\sc ii}, and S\,{\sc ii} emission lines are double-peaked. 
The double peaks are only marginally resolved in the SDSS spectrum. Reasonable fitting was achieved by using the same fitting scheme for the other lines as derived for the S\,{\sc ii} line. The velocities of the blue shifted and redshifted lines are $119\mathrm{\,km\,s}^{-1}$ and $-145\mathrm{\,km\,s}^{-1}$, respectively. 

According to \cite{Nelson}, the width of the O\,{\sc iii} line at wavelength of $5007$\AA \,($W$) can be used as a surrogate value for the velocity dispersion as $\sigma_*=W/2.35$. Therefore one can use the width of the two Gaussian profiles fitted to the O\,{\sc iii} line to estimate the masses of the two putative black holes separately in a dual AGN \citep[e.g.][]{peng}. In the case of NGC~5515, the velocity dispersions for the blue and red  lines are $\sigma_\mathrm{b}=(140.63 \pm 15.13)\,\mathrm{km\,s}^{-1}$ and $\sigma_\mathrm{r}=(131.74 \pm 7.9)\,\mathrm{km\,s}^{-1}$, respectively.
Thus using the coefficients derived for pseudobulges from \cite{Gultekin}, the assumed black hole masses are: $\log{(M^\mathrm{b}_\mathrm{BH} M^{-1}_\mathrm{\odot})}=7.29 \pm 0.5$ and $\log{(M^\mathrm{r}_\mathrm{BH} M^{-1}_\mathrm{\odot})}=7.17 \pm 0.44$, derived from the blue- and redshifted lines, respectively.
 
We compare the sum of these assumed two black hole masses with the mass estimates of \cite{Beni13b} who derived the black hole mass for NGC~5515 as a single black hole system using the $M_\mathrm{BH}-\sigma_*$ relation for galaxies containing pseudobulges \citep{Hu}, and with the scaling relation between bolometric luminosity and black hole mass \citep{VP}. The two methods yielded the following consistent values: $\log{(M^{\sigma_*}_\mathrm{BH} M^{-1}_\mathrm{\odot})}=7.45 \pm 0.25$ and $\log{(M_\mathrm{BH} M^{-1}_\mathrm{\odot})}=7.01 \pm 0.18$. Using the slightly different values for pseudobulges given by \cite{Gultekin}, a higher black hole estimate can be obtained from the $M_\mathrm{BH}-\sigma_*$ relation: $\log{(M^{\sigma_*}_\mathrm{BH} M^{-1}_\mathrm{\odot})}=7.94 \pm 0.24$.
Our sum of the assumed two black hole masses derived above ($\log{(M^\mathrm{sum}_\mathrm{BH} M^{-1}_\mathrm{\odot})}=7.54 \pm 0.48$) agrees well with the values calculated with the assumption that only one supermassive black hole resides in NGC~5515. 

Knowing the masses, we can derive the Eddington luminosities following \cite{RL}: 
\begin{equation}
L_\mathrm{Edd}=1.38 \times 10^{38} M_\mathrm{BH} M^{-1}_\odot \mathrm{\,erg\,s}^{-1}
\end{equation} 
In the scenario with two black holes separately: $L^\mathrm{b}_\mathrm{Edd}=2.7 \cdot 10^{45}\mathrm{\,erg\,s}^{-1}$ and $L^\mathrm{r}_\mathrm{Edd}=2.0 \cdot 10^{45}\mathrm{\,erg\,s}^{-1}$. If we assume that the two black holes with similar masses contribute to the measured bolometric luminosity, $L_\mathrm{bol}=(5.19\pm1.38) \cdot 10^{42}\mathrm{\,erg\,s}^{-1}$ \citep{Beni13a} evenly, we obtain an Eddington ratio of $\sim 10^{-3}$ for each. Assuming instead, a scenario where a single black hole is responsible for the observed bolometric luminosity, the implied Eddington ratio is $\sim 10^{-3} - 10^{-4}$.
All these values are within the range of Eddington ratios derived for Seyfert galaxies \citep[$10^{-4}-10^{-2}$ and $10^{-4}-10^{-1}$,][respectively]{zhang09, Sy_Singh}. Thus, based upon the mass estimates, we cannot exclude either the single or the dual black hole scenarios. 

One possible explanation for the double-peaked narrow emission lines is dual AGN, but such a spectrum can also be explained as originating from biconic outflows or rotating disks on kpc scales or otherwise disturbed NLR kinematics \citep[e.g.,][and references therein]{Liu2010, An13}. In that case, it is the same single AGN which illuminates the NLR, therefore it is expected that the blue- and redshifted lines have similar characteristics. In the case of NGC~5515, both the line widths and the line ratios (O{\sc iii}/H$\beta$, N{\sc ii}/H$\alpha$, S{\sc ii}/H$\alpha$) are very similar, equal within the uncertainties, for the blue- and redshifted components. Therefore the double-peaked emission lines can be explained in a straightforward way if both components are ionized by the same source. 

Recent studies \citep[e.g.,][]{comerford} also show that the double-peaked emission-line diagnostics alone is an inefficient way of identifying real dual AGN, but proposed that in combination with other methods such as long-slit spectroscopy and X-ray and/or radio imaging observations, dual AGN candidates can be chosen more reliably.
 
\section{Summary}
\label{sum}
Based upon the optical spectrum of NGC~5155, \cite{Beni13a} claimed that this galaxy is a good candidate for hosting dual AGN. We investigated this source using available multi-frequency radio data and new high-resolution VLBI observations conducted with the EVN. Our EVN observations revealed a compact radio emitting source with inverted spectrum, and brightness temperature exceeding $10^9$\,K. This radio emission clearly originates from a non-thermal synchrotron source associated with an AGN. The AGN nature of the radio emission is also in agreement with the long-term radio variability suggested by archival VLA data.

The position of the radio emitting source in our EVN maps agrees within the errors with the optical position of NGC~5155. We did not detect any additional radio source within $\sim 2$\,kpc of the nucleus.
According to archival VLA observations, there is no additional radio emitting feature down to a brightness level of $1 -2 \mathrm{\,mJy\,beam}^{-1}$ in the entire region covered by the galaxy in the SDSS optical image. There is an optical source at a distance of $\sim 9\arcsec$ from the radio position of NGC~5515. We did not detect any radio emission in our EVN maps at this position, either. Thus, we can exclude the possibility of having two radio-emitting AGN in the Seyfert galaxy NGC~5515. However, we cannot exclude from our VLBI data that a secondary radio-quiet AGN resides in the galaxy.

We re-analysed the SDSS spectrum of NGC~5155, and fitted the H\,$\alpha$, H\,$\beta$, O{\sc iii}, N{\sc ii}, and S{\sc ii} emission lines with double-peaked profiles. The parameters obtained from the fitting of the red and blue components are very similar. The line ratios are the same for the blue and red components. Assuming a scenario with dual AGN, we estimated the black hole masses from the blue and red components of the O{\sc iii} line, and found that the sum of the inferred two black hole masses is in agreement with the total mass estimates given by \cite{Beni13b}. 
Thus, the double-peaked narrow lines in the spectrum of NGC~5155 can be explained either by assuming two SMBHs with very similar masses and ionizing properties, from which only one is radio-emitting, or more plausibly by assuming one common ionizing source residing in the NLR, a single radio-emitting AGN.

Double-peaked narrow emission lines were originally thought to be a promising tool to select dual AGN, however their usefulness is severely questioned as other explanations (outflows, disturbed NLR kinematics, or a rotating disk) can equally well account for the observed spectral shapes in several cases.

\section*{Acknowledgments}

The EVN is a joint facility of European, Chinese, South African, and other radio astronomy institutes funded by their national research councils. The e-VLBI research infrastructure in Europe was supported by the European Community's Seventh Framework Programme (FP7/2007-2013) under grant agreement RI-261525 NEXPReS. The research leading to these results has received funding from the European Commission Seventh Framework Programme (FP/2007-2013) under grant agreement No. 283393 (RadioNet3). The National Radio Astronomy Observatory is a facility of the National Science Foundation operated under cooperative agreement by Associated Universities, Inc. For this research, K.\'E.G. was supported by the European Union and the State of Hungary, co-financed by the European Social Fund in the framework of T\'AMOP-4.2.4.A/2-11/1-2012-0001 ``National Excellence Program''. S.F. was supported by the Hungarian Scientific Research Fund (OTKA, K104539). S.F. and T.A. thank the China--Hungary Collaboration and Exchange Programme by the International Cooperation Bureau of the Chinese Academy of Sciences (CAS) for support. T.X. is supported by NSFC under Grant No. 11203056. E.K. and Z.P. acknowledge financial support from the International Space Science Institute. T.A. was supported by the 973 Program (No. 2013CB837900), NSFC (No. 11261140641), and CAS grant (No. KJZD-EW-T01). L.\'A.G. was supported by the Japan Society for the Promotion of Science. This research has made use of the NASA/IPAC Extragalactic Database (NED) which is operated by the Jet Propulsion Laboratory, California Institute of Technology, under contract with the National Aeronautics and Space Administration. Funding for SDSS-III has been provided by the Alfred P. Sloan Foundation, the Participating Institutions, the National Science Foundation, and the U.S. Department of Energy Office of Science. The SDSS-III web site is http://www.sdss3.org/.

\label{lastpage}

\end{document}